\documentstyle[12pt,epsf]{article}

\begin{document}
\pagestyle{empty}
\font\fortssbx=cmssbx10 scaled \magstep2
\hbox to \hsize{
%\special{psfile=/NextLibrary/TeX/tex/inputs/uwlogo.ps
%			      hscale=8000 vscale=8000
%			       hoffset=-12 voffset=-2}
\hskip.5in \raise.1in\hbox{\fortssbx University of Wisconsin - Madison}
\hfill$\vcenter{\hbox{\bf MAD/PH/813}
            \hbox{January 1994}}$ }
\vspace{.2in}

\begin{center}
{\LARGE Automatic Generation of Tree Level} \\
\vspace{.125in}
{\LARGE Helicity Amplitudes} \\
\vspace{.25in}
{\large T. Stelzer} \\
{\small Physics Department, University of Durham} \\
{\small Durham DH1 3LE, England} \\
\vspace{.2in}
 and \\
\vspace{.2in}
{\large W. F. Long} \\
{\small Physics Department, University of Wisconsin-Madison} \\
{\small Madison, WI 53706, USA}
\end{center}

\begin{abstract}
The program MadGraph is presented which automatically generates
postscript Feynman diagrams and Fortran code to calculate arbitrary
tree level helicity amplitudes by calling HELAS[1] subroutines.  The
program is written in Fortran and is available in Unix and VMS versions.
MadGraph currently includes standard model
interactions of QCD and QFD, but is easily modified to include
additional models such as supersymmetry.
\end{abstract}

\subsection*{1. Introduction}

Calculating tree level processes is fundamental to a large fraction of
phenomenology done today.  Although results for most interesting tree
level processes exist in the literature, it is often desirable to be able to
reproduce these calculations in order to investigate the process from
another perspective or perhaps even with just a different set of parton
distribution functions.  In principle reproducing these calculations
is trivial, in practice it can take weeks or even months to generate
and debug code for {2 $\rightarrow$ 3} or 2 $\rightarrow$ 4 processes.
Recently, several programs have been introduced to facilitate coding
tree level processes.  Some are based on symbolic manipulation
packages such as Mathmatica, Maple, and Reduce.  These packages
include HIP[2], FeynArts[3], FeynCalc[4], and Tracer[5].  They greatly aid in
calculating matrix elements using traditional trace techniques.  For
processes with more than 5 external particles, the results of trace
methods are often rather cumbersome so helicity amplitude techniques
are used.  Helicity amplitude methods sum the diagrams before
squaring, which results in the complexity growing linearly with the
number of diagrams rather than quadratically as it does using trace
techniques.  Another advantage of helicity methods is that a program
can easily be modified to decay the final state particles.  Packages
such as HELAS[1] and GRACE[6] use this technique.  While all of the above
programs are useful in the area for which they were designed, none
allow the general user to specify the initial and final state
particles to produce helicity amplitude code which the user can read
and modify or simply incorporate into a Monte-Carlo.
MadGraph was written to fulfill this function.

\pagestyle{plain}

A MadGraph user specifies
the initial state and final state particles, and the desired order in
QCD and QFD.  MadGraph then generates a Fortran function which calls HELAS
routines to calculate the amplitude.  This function can then be linked
with a generic Monte-Carlo driver with the appropriate cuts and
graphing options as well as the HELAS library.  The result is a
program which calculates the process of interest while reducing the
programming time from days to minutes.  MadGraph can generate the
Feynman diagrams and HELAS calls for hundreds of diagrams in seconds.
The color factors sometimes require a couple of minutes for
complicated QCD processes.  The code which is produced is not designed
for optimal speed, but is easy to read and debug.  As an example the
process $e^- + e^- \rightarrow e^- + e^- + Z$ runs on a DECstation
5000 with about 100 events/second.  BRS invariance and Lorentz
invariance are easy to verify using HELAS.  More elusive errors are in
overall factors.  To help minimize these errors MadGraph automatically
includes factors for the interchange of fermions and for averaging
over initial state colors.  Summing and averaging over helicity
states can either be done automatically by the MadGraph generated
function, or by the user in their Monte-Carlo.

\subsection*{2. Generating HELAS Calls}
MadGraph divides the problem of generating helicity amplitudes into
four main parts.  First, all distinct tree level topologies are
generated. Second, particles are inserted into these topologies to
produce all of the Feynman diagrams for the specified process.  Third,
the color and symmetry factors associated with each
diagram are calculated. Finally, the HELAS code for the
diagrams is generated.

The topologies are generated using a very simple recursive formula.
The program begins with the one possible topology for a process with
three external legs.  By adding an additional leg in turn to the legs
of the three topology, and to the three point vertex, the four
topologies for four external particles are generated.  The twenty five
topologies for five external particles are generated by adding one
external leg to each of the lines and three vertices of the four
particle topologies.  This process is continued to generate topologies
for six, or seven external particles.  Currently, the program is
limited to seven external particles to  limit memory use.
In principle the process can be continued to any number of particles desired.

Once the topologies have been generated, the external legs are
assigned to the particles which were requested by the user.  From
here, each vertex which has only one unspecified line is checked to
determine if the current model allows such a vertex, and if so what
particle the unspecified line must correspond to.  If more than one
particle is possible, this is noted, and the second choice is tried
later.  The process is continued until either all of the lines are
specified and the graph is stored, or a vertex is reached for which
there is no possible coupling in the specified model and the graph is
discarded.  All of the graphs which are stored are then checked to
assure they are of the correct order in the appropriate coupling
constants.  The advantage of this scheme is twofold.  First, it is
extremely fast.  Hundreds of diagrams can be generated in a fraction
of a second.  Second, it is extremely easy to add new models. All that
is necessary is to specify the allowed vertices.

The symmetry factor for the interchange of two identical fermions is
determined by following all of the fermion lines and seeing which
external particles are attached by a line.  Then one combination is
assigned to be the positive orientation. All other graphs will be
compared to this positive orientation to determine how many
permutations are required to get from one configuration to the other.
Each permutation results in a factor of -1.  The color factors are
determined by first assigning the appropriate color matrix at each
vertex, and then applying the completeness relations for the Gell-Mann
matrices as color lines are summed over.  This part of the program,
along with diagonalizing the color matrix, consumes the most time.

Generating the HELAS code for each diagram completes the process.
The design of HELAS is similar to that of the MadGraph topology algorithm.
External
wave functions are generated first and then the vertices with only one leg
left uncalculated are used to calculate the wave function for that leg.
The process is continued until all of the legs are calculated and the
last vertex gives the amplitude for the graph.  MadGraph simply looks
at the vertices and depending on what types of particles are in the
vertex, it writes out an appropriate HELAS call.  The produced code is
essentially a large `if then' block which has each possible vertex as
a case.  The code is optimized by making sure that no redundant calls
are made.  For instance, if external legs one and two are combined to
form a propagator in several different graphs, the propagator wave
function is calculated only once, and that result is then used in all
the future graphs which require it.  Although simple, this
optimization often reduces the output code by a factor of two or more.

\subsection*{3. Feynman Diagrams}
Because it is often desirable to look at Feynman diagrams for the
process being calculated, MadGraph generates a postscript file with
these diagrams and the amplitude they correspond to in the code.  In
principle, graph generation should be trivial once all of the
particles and vertices are determined.  However, determining an
esthetic arrangement of the vertices and particles requires some
thought.  The algorithm used in MadGraph minimizes the sum of the
length squared of the lines.  This does not always produce perfectly
symmetric diagrams, however it is a fairly easy procedure to implement
and the diagrams are adequate for research purposes.
For each four-gluon
vertex there are actually three identical graphs rather than just
one.  This is necessary to account for the three different color
factors in a single four gluon vertex.

\subsection*{4. Examples}
The best way to understand the program is to look at some examples.
In this section, explicit examples will show what MadGraph is
capable of performing.  A general understanding of the HELAS
package will be useful since it will allow the user to understand the
function generated by MadGraph.  If you are unfamiliar with this package, it
will suffice to
realize that given a particular set of particle momenta, HELAS returns
a complex number which is the amplitude for each Feynman diagram.
These amplitudes are then summed and squared to give the matrix
element squared.  In an effort to maintain uniformity, MadGraph asks
the same set of questions for every process it calculates.  Often
there is only one acceptable response.  If you enter a response
outside of the allowed range, or simply hit $<${\sc return}$>$, the value shown
in parenthesis is used.  One benefit of this method is that to test
the program you can enter a process and hit return for all other
requests and the default values will be used.

The information required is the process you want to have calculated,
the order of the coupling in QCD and QFD, if the Weak sector is to be
included, and a name for the process.  The process is specified by
entering the initial state particles, then a `$>$', and then the final
state particles.  The string is parsed by searching for key
characters.  Any character entered which does not represent a particle
is simply ignored.  A one digit number specifies the order of QCD from
which the order of QFD is inferred.  If the order of QFD is non zero,
and there are no external Weak bosons, you will by queried as to
whether or not to include the Weak sector.  If a `y' is found anywhere
in the response the Weak sector is included.  Finally, a name to call
the function is requested.  This name will also be used for the Fortran
file and the postscript file.  We include the
trivial example of M\rlap/oller scattering to show the format of the output.

\begin{verbatim}
   Standard Model particles include:
     Quarks:   d u s c b t d~ u~ s~ c~ b~ t~
     Leptons:  e- mu- ta- e+ mu+ ta+ ve vm vt ve~ vm~ vt~
     Bosons:   g a z w+ w- h

 Enter process you would like calculated in the form  e+ e- -> a.
 (<return> to exit MadGraph.)
e- e- -> e- e-

Attempting Process: e- e-  -> e- e-

Enter the number of QCD vertices between 0 and 0 (0):

The number of QFD vertices is 2
Would you like to include the Weak sector (n)?

Enter a name to identify process (emem_emem):

 Generating diagrams for 4  external legs
 There are  2  graphs.
 Writing Feynman graphs in file emem_emem.ps
 Reduced color matrix 1 2
 Writing function emem_emem in file emem_emem.f.

 Standard Model particles include:
     Quarks:   d u s c b t d~ u~ s~ c~ b~ t~
     Leptons:  e- mu- ta- e+ mu+ ta+ ve vm vt ve~ vm~ vt~
     Bosons:   g a z w+ w- h

 Enter process you would like calculated in the form  e+ e- -> a.
 (<return> to exit MadGraph.)

 Thank you for using MadGraph
\end{verbatim}

You can see the input lines used to specify the process as mentioned
above.  The final output lines give information about the process, the number
of graphs found, and the name of the postscript file and the function
file.  Both files are included in Appendix A.  The postscript file
contains the two diagrams for M\rlap/oller scattering.  The important
features of the Fortran file are the calling name of the two functions
and the common
block variables.  The functions have the form
\begin{verbatim}
       SEMEM_EMEM(P1,P2,P3,P4)
       EMEM_EMEM(P1,P2,P3,P4,NHEL)
\end{verbatim}
where P1...P4 are the four-momentum of particles 1 through 4
respectively.  The first function returns the amplitude summed and averaged
over helicity states, and the second
returns the amplitude for the explicit helicity state specified by {\tt NHEL},
an array of dimension four.
There are also eight common blocks
which contain information on the particle masses and the coupling
constants.  A subroutine {\tt INITIALIZE} is included in the package
which if called at the beginning will set these to reasonable values.
It is beyond the scope of this article to explain the HELAS calls, but
the HELAS manual provides a clear description of
their function.

\subsection*{5. Z + 4 Jets}

The real benefit in using MadGraph comes in evaluating processes with
many diagrams.  To illustrate this, we have chosen P P $\rightarrow$
Z + 4 jets.  There are many subprocesses possible, but the example u+g
$\rightarrow$ u+g+Z+g+g will demonstrate how to generate all of the
necessary functions.

\begin{verbatim}
   Standard Model particles include:
     Quarks:   d u s c b t d~ u~ s~ c~ b~ t~
     Leptons:  e- mu- ta- e+ mu+ ta+ ve vm vt ve~ vm~ vt~
     Bosons:   g a z w+ w- h

 Enter process you would like calculated in the form  e+ e- -> a.
 (<return> to exit MadGraph.)
ug -> ugZgg

Attempting Process: u g  -> u g z g g

Enter the number of QCD vertices between 4 and 4 (4):

The number of QFD vertices is 1
QFD required for this process ok?:
y
Enter a name to identify process (ug_ugzgg):

 Generating diagrams for 7  external legs
 There are  516  graphs.
 Writing Feynman graphs in file ug_ugzgg.ps
 Reduced color matrix 105 516
 Writing function ug_ugzgg in file ug_ugzgg.f.

 Standard Model particles include:
     Quarks:   d u s c b t d~ u~ s~ c~ b~ t~
     Leptons:  e- mu- ta- e+ mu+ ta+ ve vm vt ve~ vm~ vt~
     Bosons:   g a z w+ w- h

 Enter process you would like calculated in the form  e+ e- -> a.
 (<return> to exit MadGraph.)

 Thank you for using MadGraph
\end{verbatim}

Generating this process took about three minutes to on an
HP720.  Most of that time was spent determining the color factors,
with only a few seconds to actually draw the graphs and write the
code.  The resulting functions are
\begin{verbatim}
SUG_UGZGG(P1,P2,P3,P4,P5,P6,P7)
 UG_UGZGG(P1,P2,P3,P4,P5,P6,P7,NHEL)
\end{verbatim}
where the P's are the four-momenta of the particles and
{\tt NHEL} is a seven dimensional array which
contains the helicities of the particles.  Again there are several
common blocks which can be set up by calling the subroutine
{\tt INITIALIZE}.  However since this involves QCD, the strong coupling
constant may need to be assigned for each momentum configuration since
it depends on your choice of scale $Q^2$.  This should be done in the
subroutine which calls {\tt SUG\_UGZGG}.  The value returned
by {\tt SUG\_UGZGG} is the matrix element squared summed and averaged over
colors and helicities.  Since the helicity technique was
used, the addition of 3 lines of HELAS calls allows you to decay the Z
boson.

In order to perform a full Z + 4 jets analysis requires determining
all of the possible subprocesses listed in Table 1.  Table 1 also
shows the number of diagrams, the time needed to generate the code,
and the time needed to run the code for one set of momentum on an HP
720.

\begin{table}[htb]
\caption{Times on HP 720 in seconds}
\begin{tabular}{lrrrr}
Process                    & Diagrams& Creation& 1 Event& Optimized \\
u g $\rightarrow$ u g Z g g& 516     & 180     & 4.0    & 2.0       \\
u g $\rightarrow$ u g Z c c& 204     & 12      & 1.5    & 0.5       \\
u d $\rightarrow$ u d Z c c& 48      & 1       & 0.5    & 0.1       \\
\end{tabular}
\end{table}

The optimized speed is the speed for summing only over the non zero
helicity states.  Further speed increases can be obtained by using an
optimized color sum routine, and by using intelligent methods for
summing the helicity states.  In a fully optimized code it is
reasonable to expect about 50 events per second for the process u g
$\rightarrow$ u g Z g g.

\subsection*{6. Conclusions}
MadGraph allows the user to easily generate cross sections for complex
tree level processes along with a postscript file of the corresponding
Feynman diagrams.  All 2 $\rightarrow$ 4 processes and many 2
$\rightarrow$ 5 processes can be calculated in a nominal amount of
time.  Although the code is not fully optimized, present day
workstations are adequately powerful to integrate most processes.
Using this program will allow the physicist to concentrate on the
physics rather than on coding processes, and hopefully encourage
investigating otherwise formidable problems.
%%Madgraph is available to anyone offering me a job!......

\subsection*{Acknowledgements}
We are grateful to R.S. Fletcher and K. Hagiwara for encouragement in
pursuing this project.  We would also like to thank A. Stange and
J. Beacom for helping to check this program by comparing amplitude
calculations.  Finally we are grateful to B. Bullock for suggestions on
creating the user interface, and D. Summers for help with the
postscript routines.

\subsection*{Appendix A - Example Output Files}

MadGraph can be configured to generate either FORTRAN 77 or Fortran 90
style output functions.  Both versions of the function to compute
M\rlap/oller scattering are listed below.  The FORTRAN 77 version uses
some common language extensions: {\tt IMPLICIT NONE}, long variable names,
double precision complex,
{\tt REAL*8} and {\tt COMPLEX*16} declarations,
and the {\tt DO / ENDDO} construct.
The output file is easily edited to remove these extensions if necessary.
The Fortran 90 version conforms to the ISO Fortran standard in fixed-source
form.

\subsubsection*{emem\_emem.f, FORTRAN 77 Version}

\begin{verbatim}

      REAL*8 FUNCTION SEMEM_EMEM(P1, P2, P3, P4)
C
C FUNCTION GENERATED BY MADGRAPH
C RETURNS AMPLITUDE SQUARED SUMMED/AVG OVER COLORS
C AND HELICITIES
C FOR THE POINT IN PHASE SPACE P1,P2,P3,P4,...
C
C FOR PROCESS : e- e-  -> e- e-
C
      IMPLICIT NONE
C
C CONSTANTS
C
      INTEGER    NEXTERNAL,   NCOMB
      PARAMETER (NEXTERNAL=4, NCOMB= 16)
C
C ARGUMENTS
C
      REAL*8 P1(0:3),P2(0:3),P3(0:3),P4(0:3)
C
C LOCAL VARIABLES
C
      INTEGER NHEL(NEXTERNAL,NCOMB),NTRY
      REAL*8 T
      REAL*8 EMEM_EMEM
      INTEGER IHEL
      LOGICAL GOODHEL(NCOMB)
      DATA GOODHEL/NCOMB*.FALSE./
      DATA NTRY/0/
      DATA (NHEL(IHEL,  1),IHEL=1,4) / -1, -1, -1, -1/
      DATA (NHEL(IHEL,  2),IHEL=1,4) / -1, -1, -1,  1/
      DATA (NHEL(IHEL,  3),IHEL=1,4) / -1, -1,  1, -1/
      DATA (NHEL(IHEL,  4),IHEL=1,4) / -1, -1,  1,  1/
      DATA (NHEL(IHEL,  5),IHEL=1,4) / -1,  1, -1, -1/
      DATA (NHEL(IHEL,  6),IHEL=1,4) / -1,  1, -1,  1/
      DATA (NHEL(IHEL,  7),IHEL=1,4) / -1,  1,  1, -1/
      DATA (NHEL(IHEL,  8),IHEL=1,4) / -1,  1,  1,  1/
      DATA (NHEL(IHEL,  9),IHEL=1,4) /  1, -1, -1, -1/
      DATA (NHEL(IHEL, 10),IHEL=1,4) /  1, -1, -1,  1/
      DATA (NHEL(IHEL, 11),IHEL=1,4) /  1, -1,  1, -1/
      DATA (NHEL(IHEL, 12),IHEL=1,4) /  1, -1,  1,  1/
      DATA (NHEL(IHEL, 13),IHEL=1,4) /  1,  1, -1, -1/
      DATA (NHEL(IHEL, 14),IHEL=1,4) /  1,  1, -1,  1/
      DATA (NHEL(IHEL, 15),IHEL=1,4) /  1,  1,  1, -1/
      DATA (NHEL(IHEL, 16),IHEL=1,4) /  1,  1,  1,  1/
C ----------
C BEGIN CODE
C ----------
      SEMEM_EMEM = 0d0
      NTRY=NTRY+1
      DO IHEL=1,NCOMB
          IF (GOODHEL(IHEL) .OR. NTRY .LT. 10) THEN
             T=EMEM_EMEM(P1, P2, P3, P4,NHEL(1,IHEL))
             SEMEM_EMEM = SEMEM_EMEM + T
              IF (T .GT. 0D0 .AND. .NOT. GOODHEL(IHEL)) THEN
                  GOODHEL(IHEL)=.TRUE.
                  WRITE(*,*) IHEL,T
              ENDIF
          ENDIF
      ENDDO
      SEMEM_EMEM = SEMEM_EMEM /  4D0
      END


      REAL*8 FUNCTION EMEM_EMEM(P1, P2, P3, P4,NHEL)
C
C FUNCTION GENERATED BY MADGRAPH
C RETURNS AMPLITUDE SQUARED SUMMED/AVG OVER COLORS
C FOR THE POINT IN PHASE SPACE P1,P2,P3,P4,...
C AND HELICITY NHEL(1),NHEL(2),....
C
C FOR PROCESS : e- e-  -> e- e-
C
      IMPLICIT NONE
C
C CONSTANTS
C
      INTEGER    NGRAPHS,    NEIGEN,    NEXTERNAL
      PARAMETER (NGRAPHS=  2,NEIGEN=  1,NEXTERNAL=4)
      REAL*8     ZERO
      PARAMETER (ZERO=0D0)
C
C ARGUMENTS
C
      REAL*8 P1(0:3),P2(0:3),P3(0:3),P4(0:3)
      INTEGER NHEL(NEXTERNAL)
C
C LOCAL VARIABLES
C
      INTEGER I,J
      COMPLEX*16 ZTEMP
      REAL*8 EIGEN_VAL(NEIGEN), EIGEN_VEC(NGRAPHS,NEIGEN)
      COMPLEX*16 AMP(NGRAPHS)
      COMPLEX*16 W1(6)  , W2(6)  , W3(6)  , W4(6)  , W5(6)
      COMPLEX*16 W6(6)
C
C GLOBAL VARIABLES
C
      REAL*8         GW, GWWA, GWWZ
      COMMON /COUP1/ GW, GWWA, GWWZ
      REAL*8         GAL(2),GAU(2),GAD(2),GWF(2)
      COMMON /COUP2A/GAL,   GAU,   GAD,   GWF
      REAL*8         GZN(2),GZL(2),GZU(2),GZD(2),G1(2)
      COMMON /COUP2B/GZN,   GZL,   GZU,   GZD,   G1
      REAL*8         GWWH,GZZH,GHHH,GWWHH,GZZHH,GHHHH
      COMMON /COUP3/ GWWH,GZZH,GHHH,GWWHH,GZZHH,GHHHH
      COMPLEX*16     GCHF(2,12)
      COMMON /COUP4/ GCHF
      REAL*8         WMASS,WWIDTH,ZMASS,ZWIDTH
      COMMON /VMASS1/WMASS,WWIDTH,ZMASS,ZWIDTH
      REAL*8         AMASS,AWIDTH,HMASS,HWIDTH
      COMMON /VMASS2/AMASS,AWIDTH,HMASS,HWIDTH
      REAL*8            FMASS(12), FWIDTH(12)
      COMMON /FERMIONS/ FMASS,     FWIDTH
C
C COLOR DATA
C
      DATA EIGEN_VAL(1  )/       1.0000000000000002D+00 /
      DATA EIGEN_VEC(1  ,1  )/   7.0710678118654746D-01 /
      DATA EIGEN_VEC(2  ,1  )/  -7.0710678118654746D-01 /
C ----------
C BEGIN CODE
C ----------
      CALL IXXXXX(P1  ,FMASS(1  ),NHEL(1  ), 1,W1  )
      CALL IXXXXX(P2  ,FMASS(1  ),NHEL(2  ), 1,W2  )
      CALL OXXXXX(P3  ,FMASS(1  ),NHEL(3  ), 1,W3  )
      CALL OXXXXX(P4  ,FMASS(1  ),NHEL(4  ), 1,W4  )
      CALL JIOXXX(W2  ,W3  ,GAL,AMASS,AWIDTH,W5  )
      CALL IOVXXX(W1  ,W4  ,W5  ,GAL,AMP(1  ))
      CALL JIOXXX(W1  ,W3  ,GAL,AMASS,AWIDTH,W6  )
      CALL IOVXXX(W2  ,W4  ,W6  ,GAL,AMP(2  ))
      EMEM_EMEM = 0.D0
      DO I = 1, NEIGEN
          ZTEMP = (0.D0,0.D0)
          DO J = 1, NGRAPHS
              ZTEMP = ZTEMP + EIGEN_VEC(J,I)*AMP(J)
          ENDDO
          EMEM_EMEM =EMEM_EMEM+ZTEMP*EIGEN_VAL(I)*CONJG(ZTEMP)
      ENDDO
C      CALL GAUGECHECK(AMP,ZTEMP,EIGEN_VEC,EIGEN_VAL,NGRAPHS,NEIGEN)
      END

\end{verbatim}

\subsubsection*{emem\_emem.f, Fortran 90 Version}

\begin{verbatim}
      function semem_emem(p1,p2,p3,p4)
!
! Function generated by MADGRAPH.
! Returns amplitude squared summed/avg over colors
! and helicities
! for the point in phase space p1,p2,p3,p4,...
!
! For process : e- e-  -> e- e-
!
      implicit none
!
! Constants
!
      integer,parameter :: D = selected_real_kind(14,100)
      integer,parameter :: nexternal =     4
      integer,parameter :: ncomb     =    16
!
! Arguments
!
      real(D)                :: semem_emem
      real(D),dimension(0:3),intent(in) :: p1,p2,p3,p4
!
! External Function
!
      interface
       function emem_emem(p1,p2,p3,p4,nhel)
        real(D)                           :: emem_emem
        real(D),dimension(0:3),intent(in) :: p1,p2,p3,p4
        integer,dimension(nexternal),intent(in) :: nhel
       end function
      end interface
!
! Local Variables
!
      integer                           :: ihel,ntry
      integer,dimension(nexternal,ncomb):: nhel
      real(D)                           :: t
      logical,dimension(ncomb)          :: goodhel
!
! Helicity combination tables
!
      data goodhel/ncomb*.false./
      data ntry/0/
      data nhel(1:4,  1) /  -1, -1, -1, -1/
      data nhel(1:4,  2) /  -1, -1, -1,  1/
      data nhel(1:4,  3) /  -1, -1,  1, -1/
      data nhel(1:4,  4) /  -1, -1,  1,  1/
      data nhel(1:4,  5) /  -1,  1, -1, -1/
      data nhel(1:4,  6) /  -1,  1, -1,  1/
      data nhel(1:4,  7) /  -1,  1,  1, -1/
      data nhel(1:4,  8) /  -1,  1,  1,  1/
      data nhel(1:4,  9) /   1, -1, -1, -1/
      data nhel(1:4, 10) /   1, -1, -1,  1/
      data nhel(1:4, 11) /   1, -1,  1, -1/
      data nhel(1:4, 12) /   1, -1,  1,  1/
      data nhel(1:4, 13) /   1,  1, -1, -1/
      data nhel(1:4, 14) /   1,  1, -1,  1/
      data nhel(1:4, 15) /   1,  1,  1, -1/
      data nhel(1:4, 16) /   1,  1,  1,  1/
! ----------
! Begin Code
! ----------
      semem_emem = 0.0_D
      ntry=ntry+1
      do ihel=1,ncomb
          if (goodhel(ihel) .or. ntry < 10) then
              t=emem_emem(p1,p2,p3,p4,nhel(1:4,ihel))
              semem_emem = semem_emem + t
              if (t > 0.0_D .and. .not. goodhel(ihel)) then
                  goodhel(ihel)=.true.
                  write(*,*) ihel,t
              endif
          endif
      enddo
      semem_emem = semem_emem/ 4.0_D
      end


      function emem_emem(p1,p2,p3,p4,nhel)
!
! Function generated by MADGRAPH
! Returns amplitude squared summed/avg over colors
! for the point in phase space p1,p2,p3,p4,...
! and helicity nhel(1),nhel(2),....
!
! For process : e- e-  -> e- e-
!
      implicit none
!
! Constants
!
      integer,parameter :: D = selected_real_kind(14,100)
      integer,parameter :: ngraphs   =     2
      integer,parameter :: neigen    =     1
      integer,parameter :: nexternal =     4
      real(D),parameter :: zero = 0.0_D
!
! Arguments
!
      real(D)                :: emem_emem
      real(D),dimension(0:3),intent(in) :: p1,p2,p3,p4
      integer,dimension(nexternal),intent(in) :: nhel
!
! Local Variables
!
      integer                           :: i
      complex(D)                        :: ztemp
      real(D),dimension(neigen)         :: eigen_val
      real(D),dimension(ngraphs,neigen) :: eigen_vec
      complex(D),dimension(ngraphs)     :: amp
      complex(D),dimension(6) :: w1  , w2  , w3  , w4  , w5
      complex(D),dimension(6) :: w6
!
! Global Variables
!
      real(D)              :: gw, gwwa, gwwz
      common /coup1/          gw, gwwa, gwwz
      real(D),dimension(2) :: gal,gau,gad,gwf
      common /coup2a/         gal,gau,gad,gwf
      real(D),dimension(2) :: gzn,gzl,gzu,gzd,g1
      common /coup2b/         gzn,gzl,gzu,gzd,g1
      real(D)     :: gwwh,gzzh,ghhh,gwwhh,gzzhh,ghhhh
      common /coup3/ gwwh,gzzh,ghhh,gwwhh,gzzhh,ghhhh
      complex(D),dimension(2,12) :: gchf
      common /coup4/                gchf
      real(D)     :: wmass,wwidth,zmass,zwidth
      common /vmass1/wmass,wwidth,zmass,zwidth
      real(D)     :: amass,awidth,hmass,hwidth
      common /vmass2/amass,awidth,hmass,hwidth
      real(D),dimension(12) :: fmass,fwidth
      common /fermions/        fmass,fwidth
!
! Color Data
!
      data eigen_val(1  )/       1.0000000000000000e+00_D/
      data eigen_vec(1  ,1  )/   7.0710678118654757e-01_D/
      data eigen_vec(2  ,1  )/  -7.0710678118654757e-01_D/
! ----------
! Begin Code
! ----------
      call ixxxxx(p1  ,fmass(1  ),nhel(1  ), 1,w1  )
      call ixxxxx(p2  ,fmass(1  ),nhel(2  ), 1,w2  )
      call oxxxxx(p3  ,fmass(1  ),nhel(3  ), 1,w3  )
      call oxxxxx(p4  ,fmass(1  ),nhel(4  ), 1,w4  )
      call jioxxx(w2  ,w3  ,gal,amass,awidth,w5  )
      call iovxxx(w1  ,w4  ,w5  ,gal,amp(1  ))
      call jioxxx(w1  ,w3  ,gal,amass,awidth,w6  )
      call iovxxx(w2  ,w4  ,w6  ,gal,amp(2  ))
      emem_emem = 0.0_D
      do i = 1, neigen
          ztemp = sum(eigen_vec(:,i)*amp(:))
          emem_emem =emem_emem+ztemp*eigen_val(i)*conjg(ztemp)
      enddo
!      call gaugecheck(amp,ztemp,eigen_vec,eigen_val,ngraphs,neigen)
      end

\end{verbatim}

In addition to the Fortran source code, MadGraph produces a postscript
file containing the corresponding diagrams. For the M\rlap/oller example, the
file emem\_emem.ps prints as:

\begin{center}
\epsfxsize=4in\hspace*{0in}\epsffile{emem_emem.ps}
\end{center}

\subsection*{Appendix B - Implementation Notes}

\subsubsection*{f77 Notes}

      MadGraph is written in Fortran and uses several common
      extensions to the old FORTRAN 77 standard, including:
\begin{itemize}
      \item {\tt IMPLICIT NONE}
      \item variable names longer than 6 characters
      \item lower case letters for variable names and keywords
      \item {\tt DO/ENDDO}  and {\tt DO WHILE/ENDDO} constructs
      \item {\tt IAND} and {\tt IOR} bit intrinsics
      \item comments starting with ``!''
\end{itemize}

      In addition, f77 mode output uses the non-standard
      declarations:
\begin{itemize}
      \item {\tt REAL*8, COMPLEX*16}
\end{itemize}

      If your f77 compiler does not support {\tt IAND} and {\tt IOR}
      bit intrinsic functions (included in MIL-STD-1753)
      a C version of these functions
is supplied in the file bits.c which is replicated below:

\begin{verbatim}
	  /* bits.c */
          long iand_(long *a, long *b){return (*a & *b);}
          long  ior_(long *a, long *b){return (*a | *b);}
\end{verbatim}

\subsubsection*{f90 Notes}

      MadGraph conforms to standard[7] Fortran 90 in fixed-source
      form.  A user may convert to free-source form by adding
      trailing ``\&'' continuation characters where appropriate.

      MadGraph creates a Fortran function as part of its output.
      By default, this function is written for f77 compilers using
      the non-standard extensions noted in Appendix A.  If you prefer
      an f90 standard-conforming function, edit the beginning
      of the main program (in driver.f, or at the beginning of
      madgraph.f) to change the initialization of the variable
      {\tt fortran} to 90.

\subsubsection*{VMS Notes}

    The custom version of MadGraph for DEC VMS systems has
    modified I/O statements to compensate for differences
    between VMS and Unix formatted output.  This version
    uses some VMS-specific nonstandard extensions to
    Fortran.

\subsubsection*{Eigensystem Notes}

MadGraph requires a routine which computes the eigenvalues
and eigenvectors of a real symmetric matrix. The necessary
routines from the Lapack[8] package are included with the
distribution.  If your system already has the Lapack libraries
installed, you may remove the Lapack Fortran sources and use
your library versions.  In the single-file versions of MadGraph,
the Lapack routines are at the end of the file.  In the
tar'ed version, all the required routines are in a single file,
lapack.f.

Additional implementation information and a list of tested
platforms is included in a README file supplied with
MadGraph.

\subsection*{Appendix C - Availability}

Madgraph is available in both Unix and VMS versions {\it via}
anonymous ftp from phenom.physics.wisc.edu in the directory
pub/madgraph.  The following four files are provided:

\vspace{.2in}

\begin{tabular}{ll}
madgraph.f & Unix version as a single file. \\
madgraph.tar.Z & Unix version as a compressed tar file. \\
madgraph.tar.gz & Unix version as a gzip'ed tar file. \\
madgraph\_vms.for & VMS version as a single file. \\
\end{tabular}

\vspace{.2in}

The tar files extract into your current directory and expand to create
all the source files, a README file, and a makefile.  Edit the makefile
as appropriate for your system.

The single-file versions are provided for users without the \verb|make|
facility. These files also include commented sequence numbers in
columns 73-80.  The README file is included at the beginning as a
series of comment lines.

HEPNET users may obtain the VMS version of madgraph as a single file
by copying phenoa::local:[lib.madgraph]madgraph.for.  The executable
file for a DEC AXP (Alpha), madgraph.exe, is available in the same directory.

\subsection*{References}
\begin{description}
\item[[1]] E. Murayama, I. Watanabe and K. Hagiwara, HELAS: HELicity
Amplitude Subroutines for Feynman Diagram Evaluations, KEK Report 91-11,
January 1992.
\item[[2]] A. Hsieh and E. Yehudai, Comp. in Phys. 6 (1992) 253.
\item[[3]] J. K\"ublbeck, M. B\"ohm and A. Denner, Comp. Phys. Comm. 60 (1990)
165.
\item[[4]] R. Mertig, M. B\"ohm and A. Denner, Comp. Phys. Comm. 64 (1991) 345.
\item[[5]] M. Jamin and M. E. Lautenbacher, Comp. Phys. Comm. 74 (1993) 265.
\item[[6]] Grace Maunal: Automatic Generation of Tree Amplitudes in Standard
Models:
Version 1.0, Minami-Tateya Group (T. Ishikawa, {\it et. al.}, KEK Report 92-19,
February 1993.
\item[[7]] International Fortran Standard (ISO/IEC 1539:1991) and
American National Standard Programming Language Fortran 90 (ANSI
X3.198-1991).
\item[[8]] E. Anderson, Z. Bai, C. Bischoff, J. Demmel, J. Dongara,
J. Du Croz, A. Greenbaum, S. Hammarling, A. McKenney, S. Ostrouchov,
and D. Sorensen, Lapack Users' Guide (SIAM, Phildelphia, 1992).
\end{description}

\end{document}